\documentclass{aa}  

\usepackage{graphicx}
\usepackage{txfonts}
\begin{document}

   \title{Gas-phase formation routes of dimethyl sulfide in the interstellar medium}

    \author{
          Gabriella Di Genova\inst{1}
          \and
          Nadia Balucani\inst{1}\thanks{Corresponding author: nadia.balucani@unipg.it}
              \and
          Luca Mancini\inst{1}
          \and
          Marzio Rosi\inst{2}
          \and
          Dimitrios Skouteris \inst{3}
          \and
          Cecilia Ceccarelli \inst{4}
          }

\institute{Dipartimento di Chimica, Biologia e Biotecnologie, Università degli Studi di Perugia, Via Elce di Sotto, 8, 06123, Perugia, Italy\\
                \and    
Dipartimento di Ingegneria Civile ed Ambientale, Università degli Studi di Perugia,
 via G. Duranti, Perugia, Italy\\
                \and
                Master-Tec, Via Gerardo Dottori, 94
06132 Perugia, Italy\\
                \and
Univ. Grenoble Alpes, CNRS, Institut de Plan\'etologie et d'Astrophysique de Grenoble (IPAG), 38000 Grenoble, France\\
        }

   \date{Received April , 2025; accepted }

% \abstract{}{}{}{}{} 
% 5 {} token are mandatory
 
  \abstract
   {Dimethyl sulfide (DMS; CH$_3$SCH$_3$) is an organosulfur compound that has been suggested as a potential biosignature in exoplanetary atmospheres. In addition to its tentative detections toward the sub-Neptune planet K2-18b, DMS has been detected in the coma of the 67/P comet and toward the galactic center 
molecular cloud G+0.693-0.027. However, its formation routes have not been characterized yet.}
   {In this work, we have investigated three gas-phase reactions (CH$_3$SH +  CH$_3$OH$_2^+$, CH$_3$OH +  CH$_3$SH$_2^+$, and the CH$_3$ +  CH$_3$S radiative association)  aiming at characterizing DMS formation routes in shocked molecular clouds and star-forming regions. }
   {We have performed dedicated quantum and kinetics calculations to evaluate the reaction rate coefficients as a function of temperature to be included in astrochemical models. }
   {Among the investigated processes, the reaction between methanethiol (CH$_3$SH) and protonated methanol (CH$_3$OH$_2^+$)(possibly followed by a gentle proton transfer to ammonia) is a compelling candidate to explain the formation of DMS in the galactic center 
molecular cloud G+0.693-0.027. The CH$_3$ + CH$_3$S radiative association does not seem to be a very efficient process, with the exclusion of cold clouds, provided that the thiomethoxy radical (CH$_3$S) is available.

This work does not deal directly with the possible formation of DMS in the atmosphere of exoplanets. However, it clearly indicates that there are efficient abiotic formation routes of this interesting species.
   
   }
   {}


\titlerunning{Dimethyl sulfide formation routes}
\authorrunning{G. Di Genova et al.}

   \maketitle
%
%-------------------------------------------------------------------

\section{Introduction}
Dimethyl sulfide (DMS; CH$_3$SCH$_3$) is an organosulfur compound that has been considered as a potential biosignature in exoplanetary atmospheres since most emission sources of atmospheric DMS on Earth are associated with life \citep{DMS-terr-emission,barnes2006dimethyl,stefels2007environmental}, with the main contribution coming from microbial degradation of dimethylsulfoniopropionate (C$_5$H$_{10}$O$_5$S) \citep{stefels2007environmental}. Since there are no significant geological sources on Earth, DMS is considered to have less likely false positives than other species \citep{pilcher2003biosignatures,seager2013biosignature} and has been proposed specifically as a biosignature for identifying life on habitable exoplanets with H$_2$-rich atmospheres \citep{catling,seager2013biosignature} including the so-called "Hycean" worlds that may host habitable oceans \citep{madhusudhan2021habitability}. 
Infrared spectra recorded by the James Webb Space Telescope (JWST) NIRISS and NIRSpec instruments in the 0.9–5.2 $\mu$m range led to a tentative detection of DMS in the atmosphere of the exoplanet K2-18b, a Hycean world candidate \citep{madhusudhan2023carbon}. By using the JWST MIRI LRS instrument in the 6-12 $\mu$m range, a confirmation of that first detection was recently reported with the additional identification of other molecules, including dimethyl disulfide (DMDS), methanethiol (CH$_3$SH), phosphine (PH$_3$), and methanol (CH$_3$OH) \citep{madhusudhan2025newconstraintsdmsdmds}. 
Together with the simultaneous observation of CO$_2$ and CH$_4$ (that suggests an atmosphere far from the thermochemical redox equilibrium) and the missing detection of NH$_3$ and CO, the tentative detection of DMS drew a lot of attention for the potential indication of life.\\
\indent
Nevertheless, the assumption that DMS can only be produced by life has been challenged by its identification in astrophysical environments where biological activity is not expected. 
Indeed, the confirmed identification of DMS in the 67/P Churyumov-Gerasimenko comet by \cite{hanni2024evidence} first and then the detection of DMS in the interstellar medium (ISM) toward the Galactic center molecular cloud G+0.693-0.027 by \cite{sanz2025abiotic} clearly indicated that DMS can be formed abiotically. 
Drawing an analogy with the formation routes of dimethyl ether (DME), that is characterized by a very similar molecular structure with the central S atom substituted by an O atom, \cite{sanz2025abiotic} suggested as possible formation routes either the recombination reaction CH$_{3(ice)}$ + CH$_{3}$S$_{(ice)}$ $\rightarrow$ CH$_{3}$SCH$_{3(ice)}$ on the ice surface of interstellar dust grain or, alternatively, the gas-phase reactions  CH$_3$SH + CH$_3$SH$_2^+$ and CH$_3$OH + CH$_3$SH$_2^+$ leading to protonated DMS (that must be, then, converted into neutral DMS). 
The lack of available experimental or quantum chemistry investigations on those processes has prevented the assessment of their importance, but interesting comparisons between S and the O-related species were drawn, showing some consistency and in line with the fact that sulfur species are significantly less abundant than their O counterparts. 
We recall that O and S are supposed to have similar chemical behavior since they belong to the same group of the Periodic Table. 
However, significant differences are possible (and already noted in some cases of interest - see, for instance, \cite{SCH4} and \cite{OCH4}).\\
\indent
In this paper, we use quantum chemistry and kinetics calculations to investigate 
several possible reactions that lead to the formation of DMS or its protonated version. Since protonated methanol, CH$_3$OH$_2^+$, is much more abundant than protonated methanethiol, CH$_3$SH$_2^+$, we expect that the reaction 
\begin{equation}
    \mathrm{CH}_3\mathrm{SH} + \mathrm{CH}_3\mathrm{OH}_2^+ \rightarrow \mathrm{(CH}_3\mathrm{)}_2\mathrm{SH}^+ + \mathrm{H}_2\mathrm{O}
\end{equation}
is more relevant than the ones suggested by \cite{sanz2025abiotic}. Therefore, we have investigated this reactive system in addition to 
\begin{equation}
    \mathrm{CH}_3\mathrm{OH} + \mathrm{CH}_3\mathrm{SH}_2^+ \rightarrow \mathrm{(CH}_3\mathrm{)}_2\mathrm{SH}^+ + \mathrm{H}_2\mathrm{O}
\end{equation}
while we did not consider the reaction CH$_3$SH + CH$_3$SH$_2^+$ because of the much lower abundance of S-species compared to O-species in most extraterrestrial environments.
Both reactions (1) and (2) actually produce (CH$_{3}$)$_2$SH$^+$. 
Therefore, we have also characterized the proton transfer reaction to ammonia, following the original suggestion by \cite{charnley1995interstellar} \citep[see also][]{rodgers2001chemical,taquet2016formation} as an alternative way to dissociative electron-ion recombination for the conversion of protonated species in their neutral counterpart
\begin{equation}
    \mathrm{(CH}_3\mathrm{)}_2\mathrm{SH}^+ + \mathrm{NH}_3 \rightarrow  \mathrm{CH}_3\mathrm{SCH}_3 + \mathrm{NH}_4^+
\end{equation}
Finally, since DME can also be formed in the gas phase by the radiative association of CH$_3$ (methyl) and CH$_3$O (methoxy) radicals (as originally suggested by \cite{balucani2015} and later demonstrated by \cite{herbst}), we have also investigated the radiative association of CH$_3$ and CH$_3$S (thiomethoxy) radicals
\begin{equation}
    \mathrm{CH}_3 + \mathrm{CH}_3\mathrm{S} \rightarrow  \mathrm{CH}_3\mathrm{SCH}_3 + hv
\end{equation}
using an approach similar to the one used by \cite{herbst}.

This article is organized as follows. 
In Section 2, the employed theoretical methods are described. 
In Section 3, the results of electronic structure and kinetics calculations are presented. 
Discussion and astrochemical implications are presented in
Section 4.
The final Section 5 summarizes the principal conclusions of this work.

%--------------------------------------------------------------------
\section{Theoretical method}
In this section, we first provide details on the method employed to
obtain the stationary points of the potential energy surfaces (PESs) of the studied reactions. The description of the kinetics calculations follows.

\subsection{Electronic structure calculations}
The proposed reactions were characterized by adopting the same computational strategy already used for other bimolecular reactions (see, for instance, \cite{skouteris2015dimerization,rosi2018possible,giani2023revised}), including the related reactions  
\begin{equation}
    \mathrm{CH}_3\mathrm{OH} + \mathrm{CH}_3\mathrm{OH}_2^+ \rightarrow \mathrm{(CH}_3\mathrm{)}_2\mathrm{OH}^+ + \mathrm{H}_2\mathrm{O}
\end{equation}
\begin{equation}
    \mathrm{(CH}_3\mathrm{)}_2\mathrm{OH}^+ + \mathrm{NH}_3 \rightarrow  \mathrm{CH}_3\mathrm{OCH}_3 + \mathrm{NH}_4^+
\end{equation}
characterized in \cite{skouteris2019interstellar}.
In the case of reaction (5), the calculated rate coefficients were compared with experimental data at room temperature and higher T \citep{skouteris2019interstellar}. The excellent agreement pointed to the accuracy of our methodology for this kind of systems. 

The \textsc{gaussian09} \citep{gilbert1990theory} software package was used to investigate the PESs of reactions (1)-(4). All stationary points of the relevant PES were optimized at the DFT $\omega$B97XD level of theory \citep{chai2008systematic}, in conjunction with the minimally augmented Karlsruhe basis set ma-def2-TZVP \citep{zheng2011minimally}. Harmonic vibrational frequencies were computed, at the same level of theory, to determine the nature of each stationary point, that is, a minimum (reactants, reaction intermediates and products) if all the frequencies are real, and a first-order saddle point (transition state) if there is one, and only one, imaginary frequency. When a saddle point was found, Intrinsic Reaction Coordinate (IRC) calculations \citep{gonzalez1989improved, gonzalez1990reaction} were performed to assign each transition state to the respective reactant and product. After each geometry optimization at the DFT level, more accurate calculations were performed using the single- and double-electronic excitation coupled-cluster method with perturbative description of triple excitations, CCSD(T) \citep{bartlett1981many,olsen1996full,raghavachari1989fifth}, in conjunction with the correlation consistent valence polarized basis set \citep{kendall1992electron,woon1993gaussian,dunning1989gaussian}, augmented with a tight \textit{d} function for the sulfur atoms to correct for the core polarization effects \citep{Dunning2001} (aug-cc-pV(Q+d)Z). 
The zero-point-energy (ZPE) was used to correct all energy values.

\subsection{Kinetics calculations}
\label{kinetic}
The initial bimolecular rate constant was computed using the capture theory formalism \citep{capturetheory} according to

\begin{equation}
    k_4(E) = 2^{3/2} \pi \sqrt{\frac{C_4}{\mu}}
\end{equation}

for ion-molecule reactions (1)-(3) and

\begin{equation}
    k_6(E) = \frac{3}{2^{1/6}} \pi {\frac{C_6^{1/3} E^{1/6}}{\mu^{1/2}}}
\end{equation}

for neutral-neutral reactions. \textit{C$_n$} (with \textit{n=4} and 6 for ion-molecule and neutral-neutral reactions, respectively) is obtained by considering the entrance potential described by the formula

\begin{equation}
    V(R) = - \frac{C_n}{R^n}
\end{equation}

in which R represents the distance between the two reactants. \textit{C$_n$} is a coefficient obtained by fitting the long-range \textit{ab initio} data. 

The Ramsperger-Rice-Kassel-Marcus (RRKM) theory \citep{gilbert1990theory} was used 
to calculate the unimolecular reaction rate coefficients, $k_{uni}$, starting from the first reaction intermediates formed upon capture and as a function of the total energy, \emph{E}, of each intermediate. We have employed an in-house code widely used by us in previous studies and tested against experimental branching fractions (see, for instance, \cite{leonori2013combined,leonori2009crossed,liang,pannacci,balucani2023experimental}).
The energy-dependent rate coefficient for each specific channel $i$ is given by 

\begin{equation}
    k_{uni,i}(E) = \frac{N_{TS}(E)}{h\rho_T(E)}
\end{equation}

where \emph{N$_{TS}$(E)} is the sum of states of the transition state of each isomerization/dissociation step at the total energy \emph{E}, $\rho_T$(E) is the density of states of the intermediate undergoing the process at the total energy \emph{E}, and \emph{h} is Planck's constant.
\emph{N$_{TS}$(E)} is obtained by integrating the relevant density of states up to energy \textit{E}, assuming a rigid rotor/harmonic oscillator model. The density of states is symmetrized with respect to the number of identical configurations of the reactants and/or transition states. For the cases in which we were not able to locate a clear transition state in the exit channel, the corresponding microcanonical rate constant was obtained through a variational approach \citep{klippenstein1992variational} in which $k_{uni,i}(E)$ is evaluated at various points along the reaction coordinate. The point that minimizes the rate constant was chosen in accordance with the variational theory.  In those cases in which this approach could not be used because of problems with the electronic structure calculations,  the transition state was assumed as the products at infinite separation. This approximation has already been used by us and others and was seen to provide reasonably accurate rate coefficients \citep{pannacci,liang,vanuzzo2022reaction}. Having
obtained unimolecular rate coefficients for all intermediate steps (at a specific energy), we made a steady-state assumption for all intermediates and, by
resolving the master equation, derived energy-dependent rate
constants for the overall reaction (from initial reactants to final products).
Finally, we did Boltzmann averaging of the energy-dependent
rate constants to derive canonical rate constants (as a function of temperature).

In the case of reaction (4), the radiative association reaction rate coefficient was obtained by coupling the capture rate coefficient with the rate of spontaneous emission that we have calculated following the approach suggested by \citet{herbst1982approach}
and recently applied to the case of the related process CH$_3$ + CH$_3$O \citep{herbst}. The spontaneous emission rate $A_{E_{vib}}$ (in s$^{-1}$) can be expressed as: 

\begin{equation}
    A_{E_{vib}} = \frac{E_{vib}}{s} \sum_{i=1}^{s}\frac{A_{1\xrightarrow{}0}^{(i)}}{h\nu_i}
\end{equation}

in which \textit{s} is the number of vibrational frequencies and \textit{E$_{vib}$} is the total vibrational energy. $A_{m\xrightarrow{}n}$ is given by:
\begin{equation}
    A_{m\xrightarrow{}n} = \frac{8\pi}{c}\nu^2I
\end{equation}
where $\nu$ is the transition frequency and $I$ is the integrated intensity.

To facilitate their inclusion in astrochemical models, all the calculated rate constants have been fitted to the modified Arrhenius equation:

\begin{equation}
    K(T) = \alpha \left( \frac{T}{300}\right) ^\beta \mathrm{exp}^{-\gamma/T}.
\end{equation}

\section{Results}
\subsection{Potential energy surface of reaction CH$_3$SH + CH$_3$OH$_2^+$}
\begin{figure}[h]
    \centering
     \includegraphics[width=\hsize]{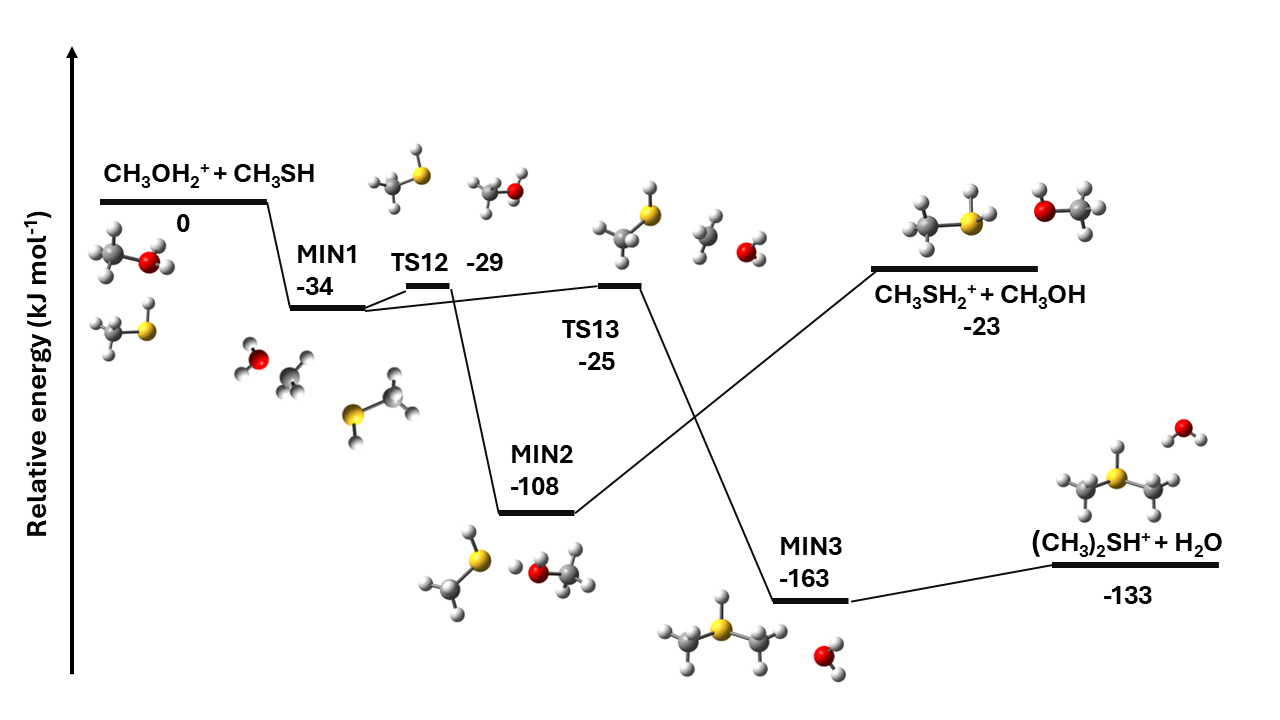}
      \caption{The potential energy surface for the reaction CH$_3$SH + CH$_3$OH$_2^+$. The indicated energy values are those computed at the CCSD(T)/aug-cc-pV(Q+d)Z level of theory.} 
    \label{fig:PES1}
\end{figure}

\begin{figure}
    \centering
     \includegraphics[width=\hsize]{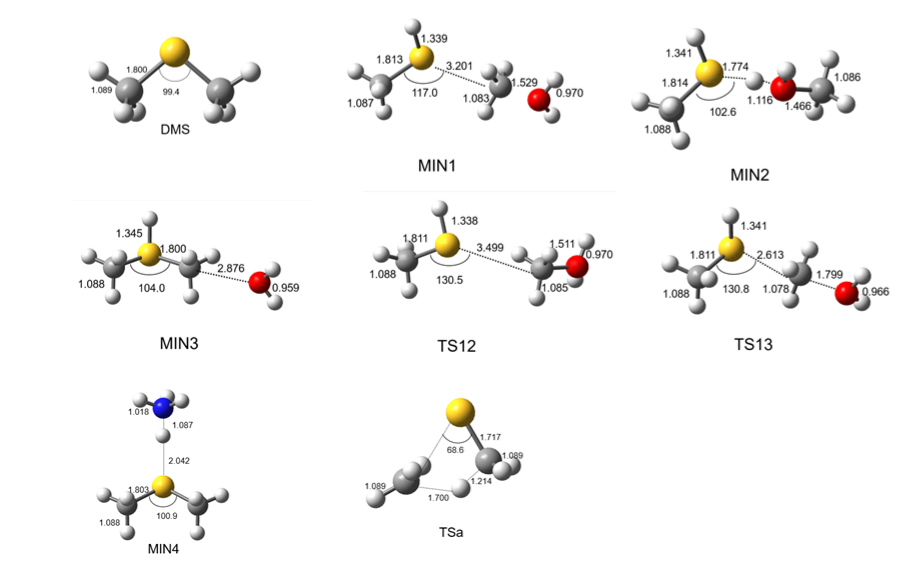}
      \caption{$\omega$B97XD/ma-def2-TZVP optimized geometries (\AA \ and °) of the intermediates and transition states identified along the PESs of reactions (1)-(4).} 
    \label{fig:stru}
\end{figure}

The PES for the reaction CH$_3$SH + CH$_3$OH$_2^+$ is shown in Fig.\ref{fig:PES1}. Each structure is optimized at the $\omega$B97XD/ma-def2-TZVP level of theory (shown in Fig. \ref{fig:stru}), while the energies indicated in Fig. 1 are those computed at the CCSD(T)/aug-cc-pV(Q+d)Z level of theory. 
The reaction between methanethiol and protonated methanol starts with a barrierless formation of the adduct MIN1 (located at - 34 kJ mol$^{-1}$ with respect to the energy content of the reactants, assumed as the zero value of the energy scale) characterized by a new C–S interaction between the sulfur atom of CH$_3$SH and the carbon atom of the protonated methanol. The C–S bond length is 3.201 \AA, which is quite larger than the typical $\sigma$ bond length, thus suggesting that this could be considered an electrostatic interaction rather than a true chemical bond. 
MIN1, through the transition state TS13, which lies at -25 kJ mol$^{-1}$ below the energy
of the reactants, can isomerize to the MIN3 intermediate (located at -163 kJ mol$^{-1}$) where the C–S
interaction becomes a true chemical bond (bond length 1.800 \AA)
while the terminal C–O bond becomes an electrostatic interaction, as we can notice from the C–O bond length, which increases from the value of 1.529 \AA \ in MIN1 to the value of 2.876 \AA \ in MIN3. Once formed, MIN3 easily evolves via the total detachment of the water molecule into (CH$_{3}$)$_2$SH$^+$. This channel is exothermic by -133 kJ mol$^{-1}$ with respect to the reactants energy asymptote.
Alternatively, 
via TS12 (located at - 29 kJ mol$^{-1}$), MIN1 can also isomerize to MIN2 (located at - 108 kJ mol$^{-1}$), another electrostatic complex where the proton of CH$_3$OH$_2^+$ is also interacting with the S atom of methanethiol. MIN2 evolves towards CH$_3$SH$_2^+$ + CH$_3$OH in a proton transfer process which is, overall, exothermic by 23 kJ mol$^{-1}$. 
In conclusion, reaction (1) is characterized by two channels: channel $(1a)$ leads to the formation of protonated DMS + H$_2$O while channel $(1b)$ is a proton transfer channel leading to protonated methanethiol and methanol. We note here that the known proton affinity of CH$_3$SH (773.4 kJ mol$^{-1}$) is larger than that of CH$_3$OH (754.3 kJ mol$^{-1}$) \citep{nistwebbook}, in agreement with our calculated enthalpy change.

In conclusion, according to our calculations, each intermediate and transition state  for reaction channel $(1a)$ lies below the reactants energy asymptote, which makes it a feasible formation route of protonated DMS in the low-temperature environments of the interstellar medium. 

\subsection{Potential energy surface of reaction CH$_3$SH$_2^+$ + CH$_3$OH}

\begin{figure}[h]
	 \includegraphics[width=\hsize]{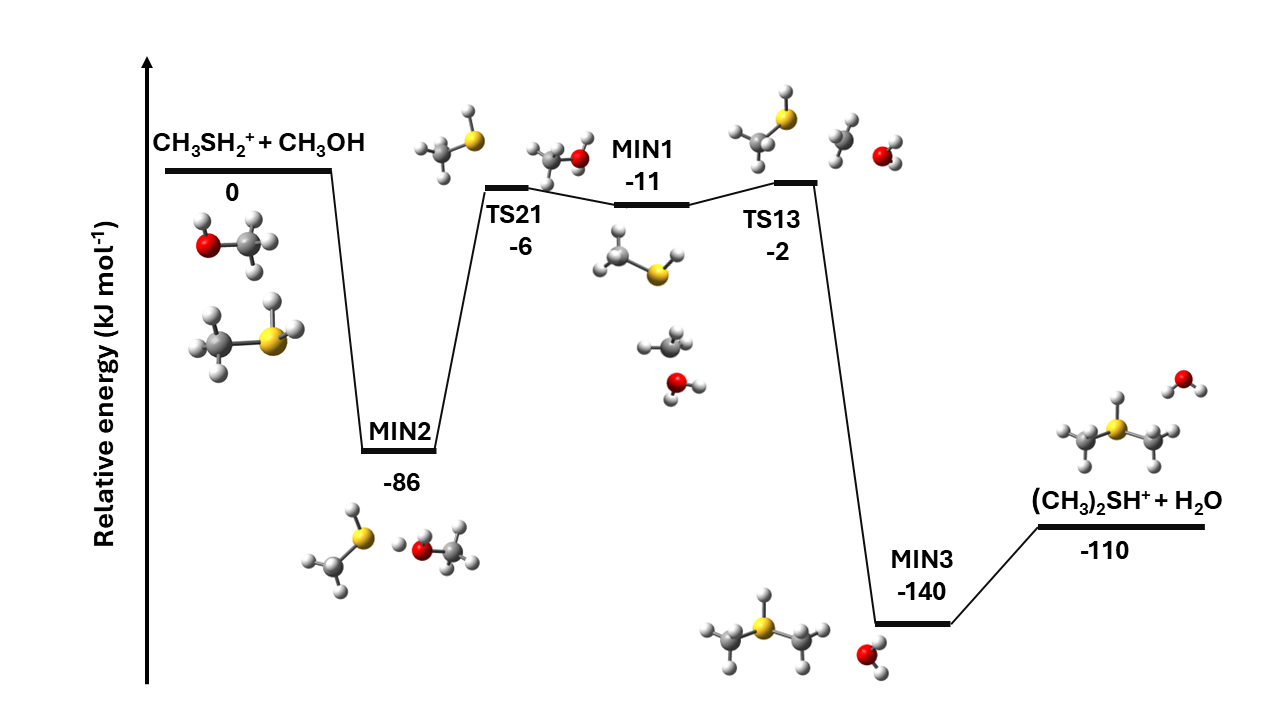}
    \caption{The potential energy surface for the reaction 
    CH$_3$SH$_2^+$ + CH$_3$OH. The indicated energy values are those computed at the CCSD(T)/aug-cc-pV(Q+d)Z level of theory.}
    \label{fig:PES2}
\end{figure}

The PES for the reaction CH$_3$SH$_2^+$ + CH$_3$OH is shown in Fig.\ref{fig:PES2}. Each structure is optimized at the $\omega$B97XD/ma-def2-TZVP level of theory (shown in Fig. \ref{fig:stru}), while the reported energies are those computed at the CCSD(T)/aug-cc-pV(Q+d)Z level of theory. 
Since here the reactants are the products of channel $(1b)$, the starting portion of this PES is equivalent to the final portion of the PES of reaction (1) in the channel $(1b)$ part. Therefore, the first step involves a barrierless formation of an electrostatic adduct (MIN2), in which the proton of CH$_3$SH$_2^+$ interacts with the oxygen atom of methanol. MIN2 lies at - 86 kJ mol$^{-1}$ below the energy asymptote of the reactants of reaction (2) that is assumed as the zero value of the energy scale for this variant. 
MIN2 can only isomerize to MIN1 (the same species as in the reaction (1) PES) by overcoming TS21 (located at - 6 kJ mol$^{-1}$). In this step, the CH$_3$SH$_2^+$ proton is being transferred to methanol, while the sulfur atom interacts with the carbon of the protonated methanol. MIN1 easily isomerizes to MIN3 via TS13 and then evolves into protonated DMS as already seen for reaction $(1a)$. In this case, there are no competitive channels. The protonated DMS + H$_2$O channel is overall exothermic by 110 kJ mol$^{-1}$ in this case. At the present level of theory, TS21 and TS13 are just below the energy of the reactants asymptote. Therefore, the energy falls within the uncertainty of the method (especially in the case of TS13). Therefore, this channel can be open or not, depending on TS21 and TS13 being submerged or not. We have performed our kinetics calculations using the values calculated at the level of theory employed. As a consequence, a strong competition from back-dissociation to reactants was seen, thus reducing the rate coefficient for the two-product exit channel. However, the rate coefficient of reaction (2) could be even lower.

\subsection{Potential energy surface of reaction (CH$_{3}$)$_2$SH$^+$ + NH$_3$}
\begin{figure}[h]
    \centering
    \includegraphics[width=\hsize]{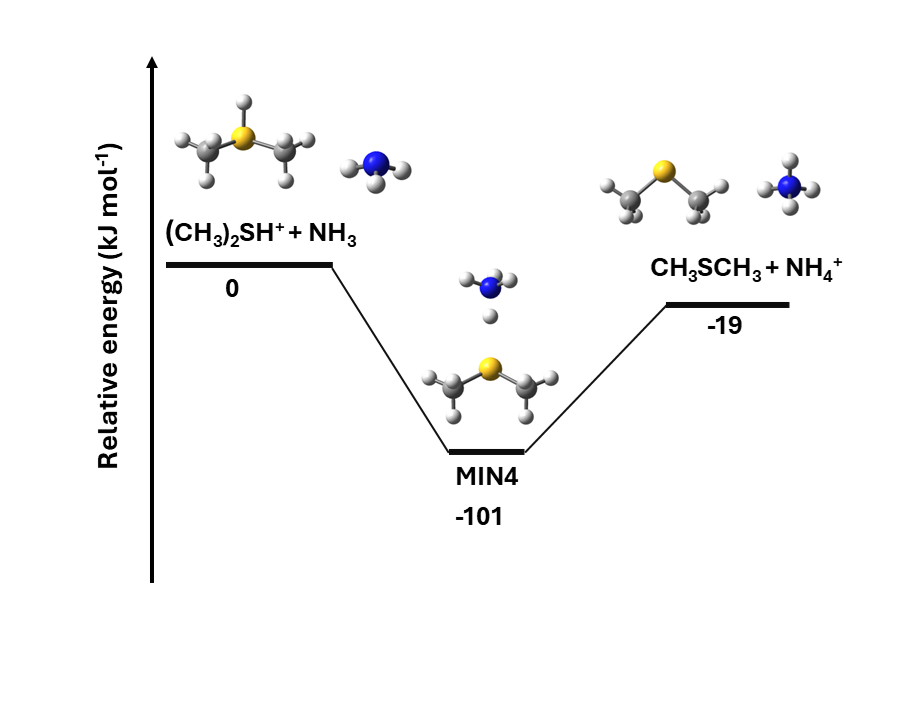}
    \caption{The potential energy surface for the reaction (CH$_3)_2$SH$^+$ + NH$_3$.The indicated energy values are those computed at the CCSD(T)/aug-cc-pV(Q+d)Z level of theory.}
    \label{fig:PES3}
\end{figure}

The PES for the reaction (CH$_{3}$)$_2$SH$^+$ + NH$_3$ is shown in Fig.\ref{fig:PES3} while the structure of the MIN4 adduct is shown in Fig. 2. Also in this case, the structures are those optimized at the $\omega$B97XD/ma-def2-TZVP level of theory, while the energies are those computed at the CCSD(T)/aug-cc-pV(Q+d)Z level of theory.
The reaction starts with the barrierless formation of an electrostatic adduct (MIN4) which lies 101 kJ mol$^{-1}$ below the reactants energy asymptote. MIN4 then evolves via a proton-transfer mechanism into DMS and protonated ammonia. The process is exothermic by 19 kJ mol$^{-1}$ with respect to the reactants energy asymptote, in line with the noted large proton affinity of CH$_3$SH. The mild exothermicity guarantees a "gentle" non-dissociative proton-transfer mechanism.

\subsection{Potential energy surface of reaction CH$_3$S + CH$_3$}

\begin{figure}
    \centering
     \includegraphics[width=\hsize]{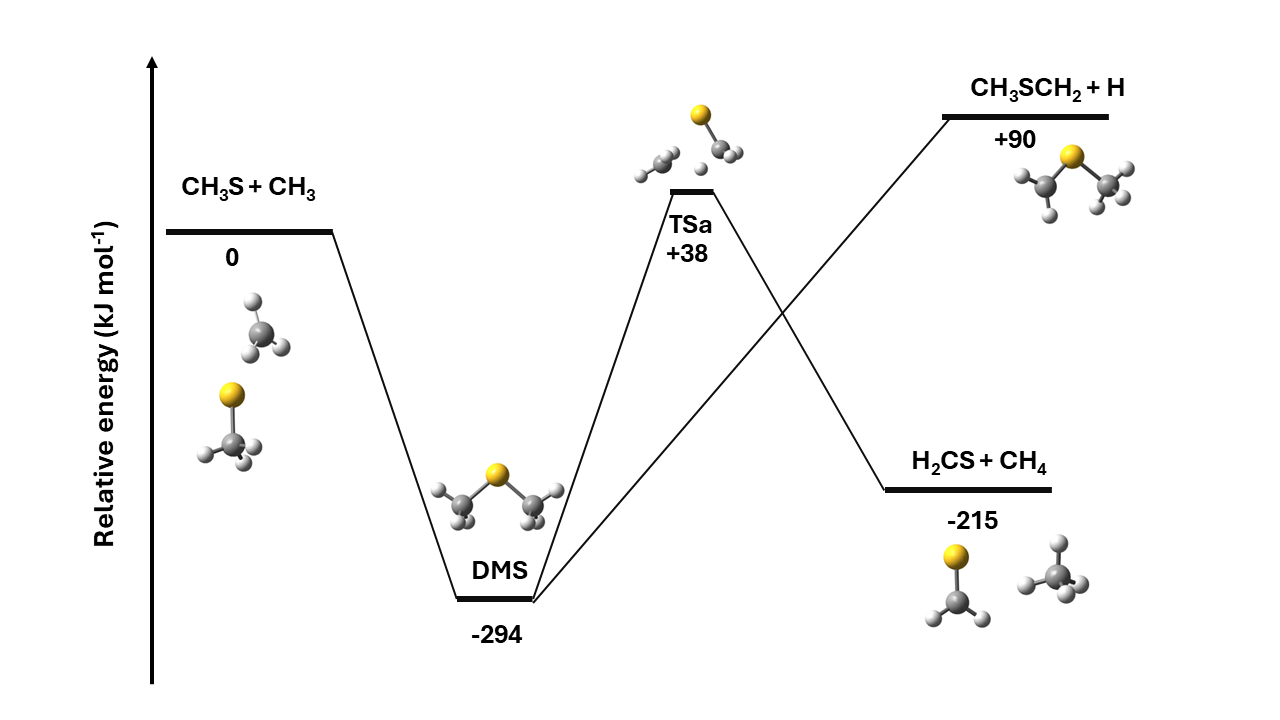}
    \caption{The potential energy surface for the reaction CH$_3$ + CH$_3$S. The indicated energy values are those computed at the CCSD(T)/aug-cc-pV(Q+d)Z level of theory.}
    \label{fig:PES4}
\end{figure}

The PES for the CH$_3$S + CH$_3$ radiative association is shown in Fig. \ref{fig:PES4}. The reported energies are those computed at the CCSD(T)/aug-cc-pV(Q+d)Z level of theory. The reaction starts with the coupling of the unpaired electrons of CH$_3$S and CH$_3$ and the formation of a new $\sigma$ bond between the carbon atom of methyl and the sulfur atom of thiomethoxy. As expected, the addition is barrierless and leads directly to the formation of DMS, located at 294  kJ mol$^{-1}$ below the reactants energy asymptote. 
In principle, DMS could evolve by losing an H atom or by forming thioformaldehyde and methane. However, the channel leading to CH$_3$SCH$_2$ + H is endothermic by 90 kJ mol$^{-1}$ and, therefore, it is not open. 
Instead, the channel leading to H$_2$CS and CH$_4$ is strongly exothermic (- 215 kJ mol$^{-1}$), but DMS must overcome an energy barrier of 332 kJ mol$^{-1}$ to dissociate into them. The related transition state (TSa) is located at + 38 kJ mol$^{-1}$ above the energy of the reactants and is therefore closed under the low T conditions typical of interstellar regions. Because of the absence of open two-product channels, radiative association can, therefore, contribute to DMS formation. 

An H-abstraction mechanism could also compete and bring directly to the 
H$_2$CS and CH$_4$ products. Normally, processes of this kind are characterized by a small entrance barrier. This was the case for the related CH$_3$O + CH$_3$ reaction \citep{SIVARAMAKRISHNAN2011618}, but here we failed to locate one with our computational methods. Therefore, we cannot exclude that this competitive process plays a role by reducing the reactive flux that will experience the presence of the deep potential well associated with DMS formation. Finally, a so-called "roaming" mechanism was seen to be active for the CH$_3$O + CH$_3$ reaction, again toward the very exothermic H$_2$CS + CH$_4$ channel \citep{SIVARAMAKRISHNAN2011618}. However, its contribution was seen to be significant only at high temperatures, which is not the case under study here. Finally, experimental and theoretical investigation of DMS thermal decomposition only provided evidence of CH$_3$ + CH$_3$S formation \citealp{themdec-exp,thermdec-theo}. 

\subsection{Rate coefficients for the CH$_3$SH + CH$_3$OH$_2^+$ reaction}
%The rate of reaction \ch{CH3SH2+} + \ch{CH3OH} as a function of the temperature is shown in Fig. \ref{fig:rate_reaction2}. 
The initial step is the formation of the MIN1 adduct. Since this is a barrierless process, we have used the approach described in Sec. 2.2 to obtain the capture rate coefficient. Then, we considered the possible evolution of MIN1 via its isomerization to MIN2 and MIN3, as well as back-dissociation to the reactants. Then we applied the Master Equation approach to derive the partial rate coefficients for back-dissociation, channel $(1a)$, and channel $(1b)$. The trend of the rate coefficients for each individual channel is shown in Fig. 5. At all temperatures, the rate coefficient of channel $(1a)$ dominates over that of the competing channel $(1b)$. With the increase of the temperature, the contribution from the back-dissociation becomes significant, thus reducing the value of $k_{(1a)}$ that, however, remains significantly large in the entire range of temperature that we have explored.

The rate constant for reaction $(1a)$ has been fitted to equation (13). To obtain a better fit, we separated the fitting considering two temperature ranges, namely 10-100 and 100-300 K. We chose to set $\gamma$=0 since the reaction is barrierless. The resulting $\alpha$ and $\beta$ coefficients, for each temperature range, are reported in Table \ref{tab:rate_coeff_react.1}. 

\begin{figure}[h]
    \centering
    \includegraphics[width=0.7\linewidth]{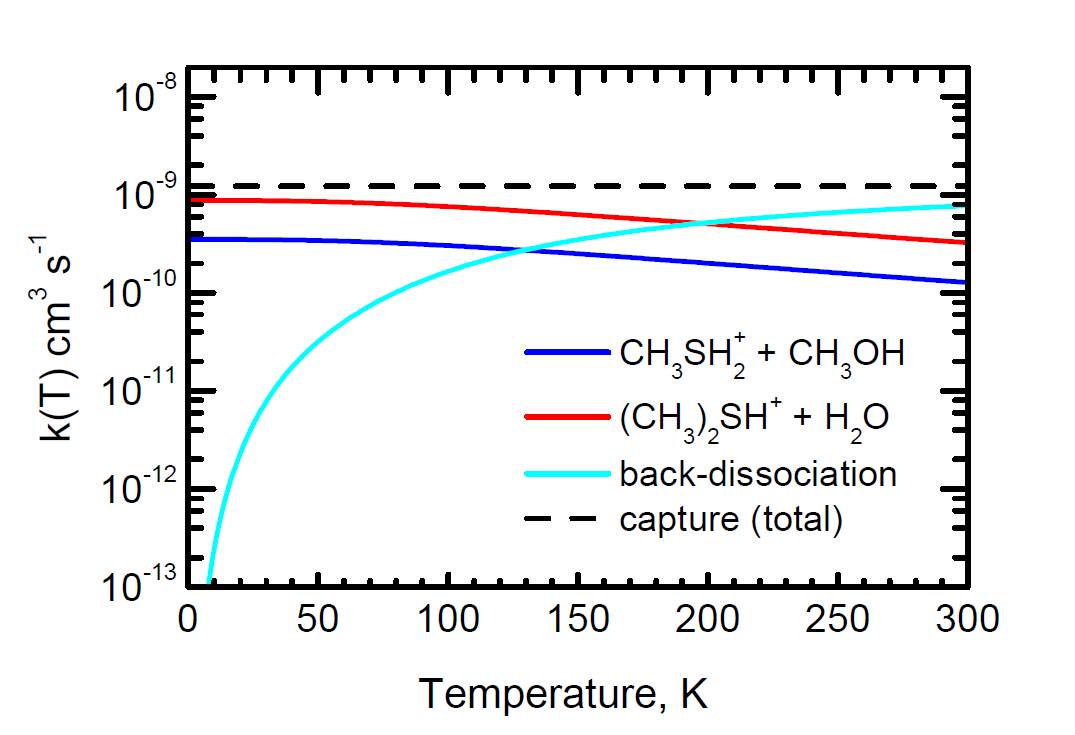}
    \caption{Rate coefficients for the reaction CH$_3$SH + CH$_3$OH$_2^+$
    as a function of the temperature. The partial rate coefficients for channel $(1a)$ leading to (CH$_3)_2$SH$^+$ + H$_2$O
     (red line), channel $(1b)$ leading to CH$_3$SH$_2^+$ + CH$_3$OH (blu line), back-dissociation (cyano line), and the global capture rate coefficient (dashed black line) are shown.} 
    \label{fig:rate_reaction2}
\end{figure}

\begin{table}[h]
\caption{Values of the $\alpha$ and $\beta$ coefficients for the CH$_3$SH + CH$_3$OH$_2^+$ $\longrightarrow$ (CH$_3)_2$SH$^+$ + H$_2$O channel of reaction (1). }
\label{tab:rate_coeff_react.1}
\centering
\begin{tabular}{c|c|c}
Temperature range (K)          &  $\alpha$ (cm$^3$s$^{-1})$ &  $\beta$                         \\ \hline
10-100 & 7.597 $\times$ 10$^{-10}$ & -0.055 \\
100-300 & 5.420 $\times$ 10$^{-10}$ & -1.647
\\ \hline
\end{tabular}
\end{table}

\subsection{Rate coefficients for the CH$_3$SH$_2^+$ + CH$_3$OH reaction}
%The rate of reaction \ch{CH3SH2+} + \ch{CH3OH} as a function of the temperature is shown in Fig. \ref{fig:rate_reaction2}. 
According to our calculations, the initial step is the formation of the MIN2 adduct that can undergo isomerization to MIN1 or dissociate back to the reactants. Since TS21 and TS13 are quite high in energy (submerged by only 6 and 2 kJ mol$^{-1}$, respectively, at this level of theory) back-dissociation will be significant.
In this case, due to some problems with the electronic structure calculations, we were not able to perform geometry optimization of the structure along the entrance channel to derive the capture rate coefficients. For this reason, a rigid scan of the entrance channel was performed, consisting of a series of single-point energy evaluations at different distances between the two interacting fragments. The resulting capture rate is significantly smaller than that for reaction (1). Once accounting for the back-dissociation, the rate coefficient for the protonated DMS formation is seen to decrease monotonically with the temperature (see in Fig. 7). As already noted, since the energy values of TS21 and, especially, of TS13 falls within the uncertainty of the method, k$_{(2)}$ could be even much smaller and neglibigle under the low T conditions of molecular clouds and star-forming regions.

The $\alpha$ and $\beta$ coefficients obtained by interpolating the calculated k$_{(2)}$, for the same temperature ranges mentioned above, are reported in Table \ref{tab:rate_coeff_react.2}. 

\begin{figure}[ht]
    \centering
    \includegraphics[width=0.7\linewidth]{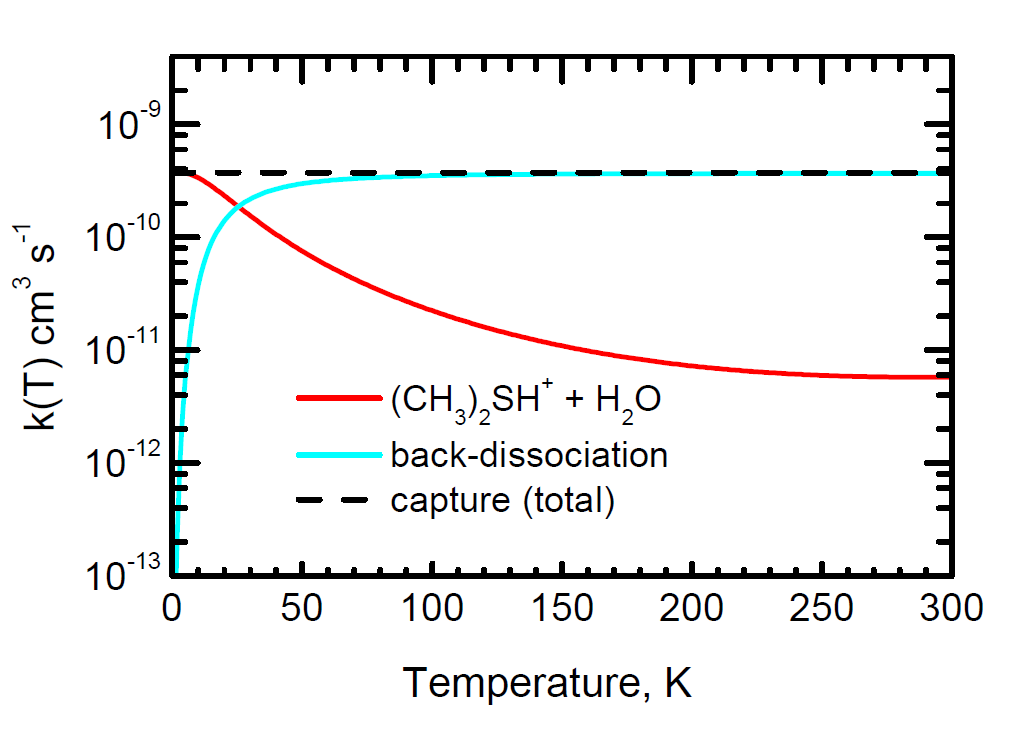}
    \caption{Rate coefficients for the reaction CH$_3$SH$_2^+$ + CH$_3$OH
    as a function of the temperature. The partial rate coefficients for the two-product reaction channel leading to (CH$_3)_2$SH$^+$ + H$_2$O
     (red line), back-dissociation (cyano line), and the global capture rate coefficient (dashed black line) are shown.}
    \label{fig:rate_2}
\end{figure}

\begin{table}[ht]
\caption{Values of the $\alpha$ and $\beta$ coefficients for the CH$_3$SH$_2^+$ + CH$_3$OH $\longrightarrow$ (CH$_3)_2$SH$^+$ + H$_2$O reaction. }
\label{tab:rate_coeff_react.2}
\centering
\begin{tabular}{c|c|c}
Temperature range (K)          &  $\alpha$ (cm$^3$s$^{-1})$ &  $\beta$                         \\ \hline
10-100 & 3.707 $\times$ 10$^{-12}$ & -1.901 \\
100-300 & 2.333 $\times$ 10$^{-12}$ & 0.143
\\ \hline
\end{tabular}
\end{table}

\subsection{Rate coefficients for the CH$_3$ + CH$_3$S radiative association}
\begin{figure}[ht]
    \centering
    \includegraphics[width=0.7\linewidth]{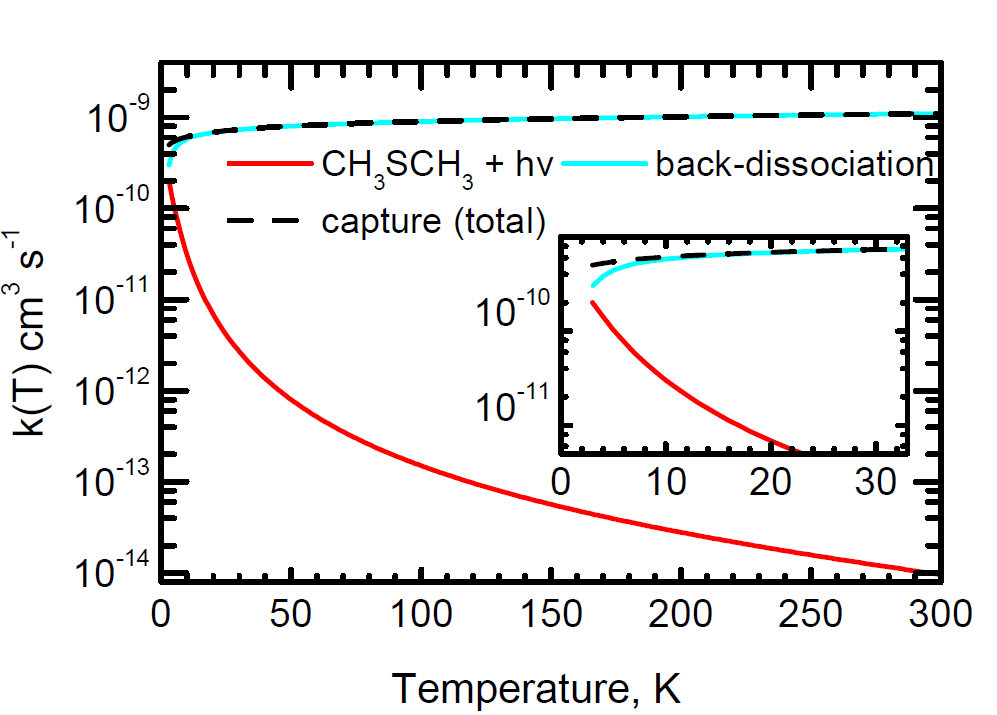}
    \caption{Rate coefficients for CH$_3$ + CH$_3$S radiative association reaction (red line) and back-dissociation (cyano lines). The global capture rate coefficients are also shown (black dashed line). A close-up for the range of temperatures between 5 and 30 K is shown in the inset figure.}
    \label{fig:rate_RA_DMS}
\end{figure}

The capture rate coefficients and their partitioning among radiative association and back-dissociation are reported for the CH$_3$ + CH$_3$S system in Fig. \ref{fig:rate_RA_DMS}.
Because of the competition with the back-dissociation, the process of radiative association can give a significant contribution only at very low temperatures. The rate constant for reaction 3 has been fitted to equation (3). The reaction rate coefficients obtained from the fitting are reported in table \ref{tab:rate_coeff_react.3}.

\begin{table}[ht]
\caption{Values of the $\alpha$ and $\beta$ coefficients for reaction (3), namely CH$_3$S
 + CH$_3$ $\longrightarrow$
  CH$_3$SCH$_3$. }
\label{tab:rate_coeff_react.3}
\centering
\begin{tabular}{c|c|c}
Temperature range (K)          &  $\alpha$ (cm$^3$s$^{-1})$ &  $\beta$                         \\ \hline
10-100 & 9.630 $\times$ 10$^{-15}$ & -2.524 \\
100-300 & 1.107 $\times$ 10$^{-14}$ & -2.653
\\ \hline
\end{tabular}
\end{table}

\section{Discussion}

\subsection{Comparison of the reactions involving S- and O-bearing species}
%Let's begin the discussion by comparing the reactions we have studied here with analogous reactions involving oxygen species. The CH$_3$OH + CH$_3$OH$_2^+$ reaction has been the focus of both theoretical and experimental investigations (see \cite{skouteris2019interstellar} and references therein).

A first interesting point to discuss is the comparison of the reactions studied in the present work with the analogous reactions involving oxygen species. 
The CH$_3$OH + CH$_3$OH$_2^+$ reaction has been the focus of both theoretical and experimental investigations \citep[see][and references therein]{skouteris2019interstellar}.
Notably, our research group conducted a recent theoretical study employing the same methodology used here. 
When we compare the PES of the CH$_3$OH + CH$_3$OH$_2^+$ system with that of reaction (1), we can identify both similarities and differences. 
The initial approach and the subsequent steps bear a close resemblance; however, some distinctions arise. 
First, the CH$_3$OH + CH$_3$OH$_2^+$ reaction is symmetrical, which is not the case for the systems addressed in this work. 
In this sense, the equivalent reaction would be CH$_3$SH + CH$_3$SH$_2^+$; however, we opted not to study it since methanol (and by extension, protonated methanol) is significantly more abundant than methanethiol. 
Consequently, the symmetrical reaction is much less probable under interstellar medium conditions than the reaction involving either protonated methanol or methanol. 
We recall that \cite{taquet2016formation} demonstrated the role of protonated methanol as a donor of a methyl group in many reactions leading to interstellar complex organic molecules. 
The symmetry of the CH$_3$OH + CH$_3$OH$_2^+$ system has the effect of rendering the formation of MIN2 (an intermediate of that PES with a structure similar to the one of MIN2 of Fig. 2, see Fig.3 and 4 of \cite{skouteris2019interstellar}) a dead end, as it can only revert to the starting reactants. 
Conversely, in reaction (1), MIN2 leads to the competing proton transfer process. 
Thus, the first notable difference is that in reaction (1), there are two competing reaction pathways. 
Additionally, due to the different characteristics of MIN3, which has a sulfur atom at the center of the structure rather than an oxygen atom, we lack an equivalent pathway that proceeds via another intermediate (MIN5 in \cite{skouteris2019interstellar}, the ultimate fate of which is the formation of protonated DME and water). 
Another significant difference regards the exothermicity of channel $(1a)$, which is significantly lower (-133 kJ mol$^{-1}$ compared to -62.8 kJ mol$^{-1}$ for the CH$_3$OH + CH$_3$OH$_2^+$ $\rightarrow$ (CH$_3)_2$OH$^+$ + H$_2$O reaction). 
The other minima and transition states of the two PESs are substantially similar, both for the molecular structure and energy position. 
Overall, the capture rate coefficients for reactions (1) and (5) are quite similar, but the presence of the competing pathway $(1b)$ slightly reduces the $k_{(1a)}$ value. 
In both cases, with the increase of the temperature, back-dissociation becomes significant. 

Concerning reaction (3), there is also a strong similarity with the equivalent (CH$_3)_2$OH$^+$ + NH$_3$ reaction \citep{skouteris2019interstellar}. 
Since the proton affinity of DMS is larger than that of DME (830.9 vs 792 kJ mol$^{-1}$) \citep{nistwebbook}, the proton transfer reaction with ammonia is less exothermic. 
The resulting rate coefficient is very similar in both cases.

Finally, concerning the radiative association process, we note that there are significant differences with respect to the CH$_3$ + CH$_3$O case. 
The canonical rate coefficient calculated by \cite{herbst} is significantly larger than ours for temperatures larger than $\sim10$ K. 
Their phase-space coefficient is, instead, smaller than our values for very low temperatures, but it remains in the 2 $\times 10^{-12}$ cm$^3$ s$^{-1}$ range also at 300 K. 
Globally, our rate coefficient for the CH$_3$ + CH$_3$S radiative association is around 2 orders of magnitude smaller with respect to the case of CH$_3$ + CH$_3$O. 
Trying to rationalize such a difference, we note that the potential energy well associated with DMS is -234 kJ/mol with respect to CH$_3$ + CH$_3$S, while for DME it is -351 kJ/mol. 
This should increase the DME lifetime, but the density of states is larger in the case of DMS and, therefore, these two factors contrast each other. 
At the same time, the spacing of the vibrational levels is larger in the case of DME, and this has the effect of increasing the Einstein coefficient for spontaneous emission, which varies with the square of the frequency. 
Using the data of Table 1 in \cite{herbst} we calculated the Einstein coefficients for DME and compared them with ours. 
Around 300 K, they are larger by a factor of 3. 
Therefore, our conclusion is that it is the different capability of emitting photons that makes the difference. 
In all cases, even though the radiative association has a much smaller rate coefficient, we note that its contribution can play some role up to 100 K in very rarefied media, where a few other processes can provide a contribution.

\subsection{Astrophysical implications: the case of molecular clouds and shocked regions}

To date, the only detection of DMS in the interstellar medium has been toward the galactic center molecular cloud G+0.693-0.027 by \cite{sanz2024interstellar}.
This object is known to be less depleted of S-species in the gas phase with respect to most known star-forming regions and molecular clouds \citep{sanz2024interstellar}, where S is believed to be mostly locked in the dust grains \citep{fuente2023gas}. 
The reduced depletion in G+0.693-0.027 is probably associated with the erosion of the dust ices driven by shocks \citep{requena2006organic,zeng2020cloud}. According to the estimates by \citep{sanz2025abiotic}, the ratios between the abundances of DME and DMS is $\frac{[DME]}{[DMS]}=30 \pm 4$ and that between CH$_3$OH and CH$_3$SH is $\frac{[CH_3OH]}{[CH_3SH]} \approx 31$. 
Considering a temperature of 100 K and a kinetically-controlled regime in which the destruction routes of both DME and DMS have similar rate coefficients and $k_5 \approx k_{(1a)}$, if reaction $(1a)$ is the main DMS formation routes and reaction (5) is the main DME formation route, and considering that the proton transfer reactions with ammonia will have the same rate coefficient for protonated DMS and DME, we can write

\begin{equation}
\begin{split}
    \frac{[DME]}{[DMS]} \propto  \frac{rate_{DME_{formation}}}{rate_{DMS_{formation}}} =\ \ \ \ \ \ \ \ \ \ \ \ \ \ \ \ \ \ \ \ \ \ \ \ \ \ \ \\ \frac{k_5 \times [CH_3OH] \times [CH_3OH_2^+]}{k_{(1a)} \times [CH_3SH] \times [CH_3OH_2^+]} = 
    \frac{k_5 \times [CH_3OH] }{k_{(1a)} \times [CH_3SH]} \approx  \frac{[CH_3OH]}{[CH_3SH]}
     \end{split}
\end{equation}
which is nicely in agreement with the observed abundance ratios reported by \cite{sanz2025abiotic}.
As a matter of fact, the value of $k_5$ is around 5 $\times$ 10$^{-10}$ cm$^3$s$^{-1}$ while the value of $k_{(1a)}$ round 7 $\times$ 10$^{-10}$ cm$^3$s$^{-1}$. 
However, given the uncertainty of the detection and, especially, the missing information on the destruction routes, we can conclude that the agreement is quite satisfactory and that CH$_3$OH and CH$_3$SH can reasonably be the parent molecules of DME and DMS via their reactions with protonated methanol, followed by proton transfer to ammonia.

We plan to run a complete astrochemical model to verify this very simplified estimate in a forthcoming article.

\subsection{Astrophysical implications for other environments}

\cite{hanni2024evidence} has confirmed the detection of DMS in the comet 67/P Churyumov-Gerasimenko initially reported with some ambiguity because of the presence of ethanethiol, the DMS isomer \citep{altwegg2019cometary}. 
Molecules detected in cometary comae could be primary, that is, released directly by the nucleus of the comet, or secondary, that is, formed by chemical processes in the gaseous envelope formed by the sublimation of parent species. 
\cite{hanni2024evidence} did not discuss the profile of DMS with respect to the distance from the nucleus, and, therefore, we do not know to which class it belongs. 
Here, we comment that the ion CH$_3$OH$_2^+$ has been clearly detected by ROSINA \citep{beth} as well as methanethiol \citep{calmonte2016sulphur}. 
Therefore, at least in principle, reaction (1) can give a contribution to DMS formation in the gaseous coma.

Concerning the DMS observed in the planet K2-18, we are not aware that ion-molecule reactions are considered in the photochemical models of the atmosphere of sub-Neptune exoplanets. 
Our work, therefore, can only contribute by pointing out that the CH$_3$ + CH$_3$S recombination is possible, either in high-pressure regions via termolecular collisions or by radiative association in the uppermost rarefied atmosphere. 
The presence of the thiomethoxy radical would require, as a precursor species, methanethiol  which is also considered a possible biosignature candidate, though less strong than DMS or dimethyl disulfide.

\section{Conclusions}
In this work, we have reported new formation routes of DMS based on dedicated quantum and kinetics calculations. 
Our aim was to characterize the formation of DMS in shocked molecular clouds or star-forming regions. 
We have identified three possible processes. 
Among them, we believe that the reaction between methanethiol and protonated methanol is a compelling candidate to explain the formation of DMS in the galactic center molecular cloud G+0.693-0.027. 
The CH$_3$ + CH$_3$S radiative association does not seem to be a very efficient process, with the exclusion of cold clouds, provided that the thiomethoxy radical is available.

This work does not deal directly with the possible formation of DMS in the atmosphere of exoplanets. 
However, it clearly indicates that there are efficient abiotic formation routes of this interesting species.

\begin{acknowledgements}
The authors thank the Italian MUR PRIN2020 (2020AFB3FX-Astrochemistry beyond the second period elements) and the European Union’s Horizon 2020 research and innovation program from the European Research Council (ERC) (project ‘The Dawn of Organic Chemistry’ DOC, grant agreement No 741002) for support.
Support from the Italian Space Agency (Bando ASI Prot. n. DC-DSR-UVS-2022-231, Grant no. 2023-10-U.0 MIGLIORA) is also acknowledged.
MR acknowledges financial support under the National Recovery and Resilience Plan (NRRP), Mission 4, Component 2, Investment 1.1, Call for tender No. 104 published on 2.2.2022 by the Italian Ministry of University and Research (MUR), funded by the European Union – NextGenerationEU– Project Title 2022JC2Y93 ChemicalOrigins: linking the fossil composition of the Solar System with the chemistry of protoplanetary disks – CUP J53D23001600006 - Grant Assignment Decree No. 962 adopted on 30.06.2023 by the Italian Ministry of Ministry of University and Research (MUR).

\end{acknowledgements}

%\begin{thebibliography}{}

%    \bibitem[1988]{balluch} Balluch, M. 1988,
%      A\&A, 200, 58

%    \bibitem[1980]{mizuno} Mizuno H. 1980,
%      Prog. Theor. Phys., 64, 544
   
%   \bibitem[1987]{tscharnuter} Tscharnuter W. M. 1987,
%      A\&A, 188, 55
  
%   \bibitem[1980a]{yorke80a} Yorke, H. W. 1980a,
%      A\&A, 86, 286

%   \bibitem[1997]{zheng} Zheng, W., Davidsen, A. F., Tytler, D. \& Kriss, G. A.
%      1997, preprint
% \end{thebibliography}
%

\end{document}